# Sigma-phase in Fe-Cr and Fe-V alloy systems and its physical properties

Stanisław M. Dubiel[*] and Jakub Cieślak

Faculty of Physics and Applied Computer Science, AGH University of Science and Technology, 30-059 Kraków, Poland

## Abstract

A review is presented on physical properties of the σ-phase in Fe-Cr and Fe-V alloy systems as revealed both with experimental – mostly with the Mössbauer spectroscopy - and theoretical methods. In particular, the following questions relevant to the issue have been addressed: identification of σ and determination of its structural properties, kinetics of α-to-σ and σ-to-α phase transformations, Debye temperature and Fe-partial phonon density of states, Curie temperature and magnetization, hyperfine fields, isomer shifts and electric field gradients.

## Table of contents

## 1. INTRODUCTION

Among the complex so-called Frank-Kasper phases that can be geometrically described in terms of the basic coordination polyhedron with coordination number (CN) equal to 12, 14, 15 and 16 [1], the sigma phase that can occur in transition-metal alloy systems, is known as the one without definite stoichiometric composition. Consequently, it can exist in different composition ranges and, therefore, its physical properties can be tailored by changing its compositional elements and, within a given constitution, by changing chemical composition. Among about 50 examples of binary σ phases that have been reported so far in the literature, only two viz. σ-FeCr and σ-FeV are known to have well evidenced magnetic properties [2-8]. From the two, the σ-FeCr is regarded as the archetype, because it was the Fe-Cr system in which the presence of the σ phase was suggested as early as 1907 for the first time to occur [9]. However, its real discovery came only 20 years later when Bain and Griffits [10] and independently Chevenard [11], found in a Fe-Cr-Ni alloy a hard, very brittle and non-magnetic phase which they termed as the "B constituent". The name "sigma" was given by Jette and Foote [12], and the definite identification of its crystallographic structure as close-packed tetragonal – space group $D^{14}_{4h}$ - $P4_2/mnm$ – with 30 atoms per unit cell was done Bergman and Shoemaker [13]. The atoms were revealed to be distributed over five crystallographically non-equivalent lattice sites: A(I), B(II), C(III), D(IV) and E(V) with occupation numbers 2 (a), 4 (f), 8 (i), 8 (i´) and 8 (j), respectively (with Wyckoff letter in the bracket). The sites have rather high coordination numbers – CN=12 for sites A and D, CN=14 for sites C and E, and CN=15 for site B – and quite different local symmetries. For several years it was believed that Fe atoms occupied exclusively the sites A and C, Cr atoms were solely situated on sites C, while the both kinds of atoms could be found on sites D and E. Such situation, from the viewpoint of the Mössbauer Spectroscopy (MS), resulted in an oversimplified interpretation of Mössbauer spectra i.e. instead of five subspectra, three to four ones were fitted [14-17]. Finally, it has been proved that the occupation of all sites is mixed i.e. they are populated by both kinds of elements constituting a given alloy. However, their distribution was revealed to be not random; A and D sites are preferentially occupied by Fe atoms while Cr and V atoms prefer to sit on B, C and E sites [18,19]. That means that each spectrum is composed of at least five subspectra. From this knowledge, one cannot, however, make an easy use when analyzing the Mössbauer spectra due to their low spectral resolution following from weak magnetic properties, a complex non-stoichiometric structure and a lower than the cubic symmetry. As the magnetic interactions are comparable with quadrupole ones, the spectra analysis is a difficult task. Sample preparation conditions have quite moderate effect on the site-occupation numbers in the Fe-Cr system: the most effective being the cold rolling, for which the overall change of the site-occupancy was equal to ~19%, and the least effective was the time of annealing for non-deformed samples, where the elongation of the annealing time from 35 to 277 days resulted in an overall change of the site-occupancy merely by ~10%. Particular sites are not equivalent as far as the sample preparation conditions are concerned: the least influenced was site D followed by E, and the most influenced sites were sites B and A [19].

Another reason that the σ-FeCr has been of a special interest follows from the fact it may precipitate in stainless steels (SS). Since about a century SS have been a major construction material in many important branches of industry like: chemical and petrochemical, conventional and nuclear power plants just to mention a few. Frequently, equipment manufactured from SS is in service at elevated temperatures due to its exceptional resistance to a high-temperature corrosion. However, at such temperatures SS is thermodynamically unstable and eventually it either decomposes into Fe-rich (α) and Cr-rich (α') phases, causing the so-called "475°C embrittlement", or it transforms partly into σ. The presence of the latter, even at the level of a few percent, results in a drastic deterioration of many useful properties [3]. Hence, from a practical viewpoint, the σ-phase is an unwanted phenomenon. However, from the scientific point of view, it is very interesting not only *per se* as a very complex phase, hence a challenging one, but also as it can be treated as a very good reference system for investigation of physical properties of amorphous solids and glasses [20].

Several experimental techniques were successfully applied to study the σ-phase in Fe-Cr and Fe-V systems, among them MS has played an important role. One of the obvious reasons for that position of MS follows from the fact that the Mössbauer element viz. Fe is the main constituent of these alloy systems which makes MS applicable to study the issue. However, MS offers much more viz. it can be quantitatively used to: (a) identify σ, (b) study the kinetics of α-to-σ transformation and the reverse one, (c) determine the Debye temperature, the Curie temperature, the magnetic hyperfine field (spin-density) and isomer shift (charge-density).
Consequently, the main body of the present paper is devoted to the kinetics of formation and to physical properties, such as dynamic and magnetic ones, of the σ-phase in the Fe-Cr and Fe-V alloy systems.

## 2. RANGES OF OCCURANCE OF σ IN Fe-Cr AND Fe-V SYSTEMS

### 2. 1. Fe-Cr

Several investigators have contributed to determination of the range of occurrence of σ in the bulk alloy system and its actual borders in terms of characteristic temperatures and composition were changed in the course of time.
Many investigators were involved in finding the high-temperature limit for the formation of σ. The values of this temperature for the bulk alloy differ by ~100$^o$C. Several authors gave ~820$^o$ C as the lowest value of the upper-limit of the existence of σ as a single phase [21-23]. Slightly higher value i.e. ~830$^o$C was reported by Kubaschewski [24], while a much higher one viz. ~930$^o$C was claimed in [25].
Williams and Paxton [26], and Williams [27] proposed a significant modification of this phase diagram by postulating an existence of a low-temperature limit estimated at ~520$^o$C. At this temperature, the σ-phase should undergo a eutectoid transformation (phase decomposition) into α and α' phases. The latter found a support in form of theoretical calculations by Kubaschewski and Chart, who, however, determined the eutectoid decomposition temperature as 460$^o$C [28], which seems to be the lowest value for this characteristic temperature, while the highest one of ~650$^o$C was reported by Pomey and Bastien [29]. There are also relevant data lying between the two extreme values like: ~630$^o$C [30], ~540-560$^o$C [31], or < 480$^o$C [32].
Another aspect of the issue depicts a compositional range in which σ as a single-phase can be formed. Also here there is a lack of agreement i.e. various authors give different values for the range (in at% Cr): 41- 51 [31], 42-51 [29], 43.5 -50 [32], 44-50 [22, 33].
The difficulty in accurately determining the phase boundaries in terms of the temperature and composition follows from several factors like:
  (a) Slowness of the transformation
  (b) Dependence on purity of alloys
  (c) Dependence on mechanical state (strained or strain-free state)
  (d) Dependence on the temperature of annealing
Consequently, the composition and temperature ranges of the single σ-phase existence as found in various studies are very different.
The "official" phase diagram of the Fe-Cr alloy system showing, in particular, the borders of the σ-phase occurrence is presented in Fig. 2.1.
The existence of σ in Fe-Cr was predicted theoretically and corresponding phase diagram was calculated, too [34, 35].
It should also be mentioned that thin films of quasi-equiatomic Fe-Cr alloys show different phase boundary for the σ-phase than the bulk alloys [37]. In the alloy containing 45 wt% Cr this phase was found to exist between 300 and 1025°C. According to [38], the low temperature boundary lies even lower viz. between 200 and 300°C.

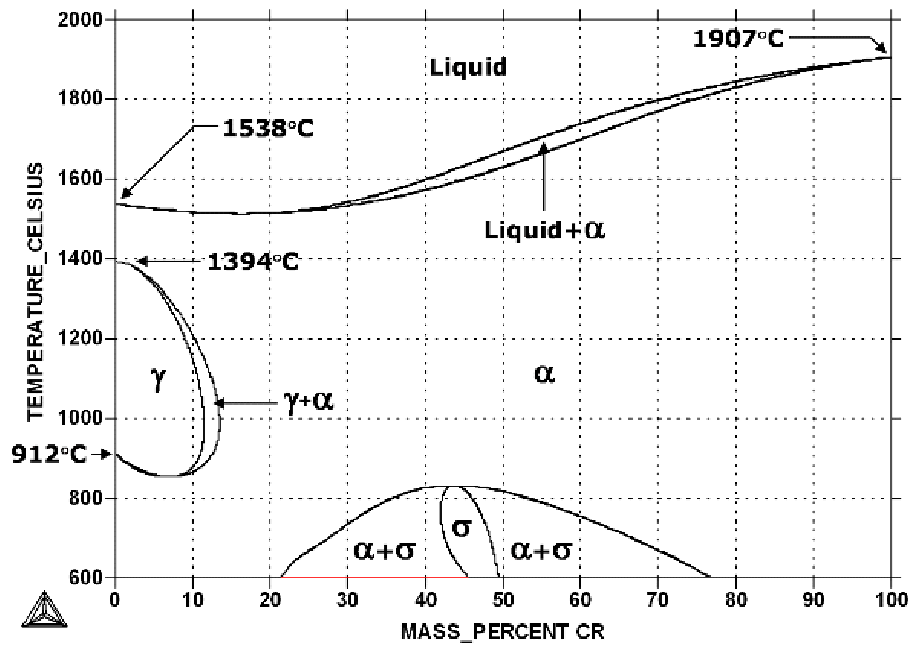

**Fig. 2.1**
*Phase diagram of a bulk Fe-Cr alloy system [36].*

**2. 2. Fe-V**

Despite the σ-phase in the Fe-V system was found many years ago [10, 21], the concentration and temperature ranges of its existence are still not precisely known. Reasons for this situation are similar to those mentioned in paragraph 2.1.
The probability of occurrence of the σ-phase in this alloy system is about 20 times larger than that in Fe-Cr, which follows both from a wider compositional and temperature ranges. However, also here there are differences with regard to the values of borders in which σ as the single-phase can be formed. They depict both temperature as well as composition.
According to one of the "official" phase diagrams, which is presented in Fig. 2.2, the upper temperature limit for the formation of σ is equal to 1321°C for V content of x = 46.5 wt%, and it rather strongly depends on composition e. g. ~1000°C for x = 34 at% and ~1050°C for x = 63 at% V [21]. Much lower value for the maximum temperature viz. ~1224°C can be found elsewhere [21,25]. On the other hand, similar values are reported for the lower limit temperature e.g. ~647°C [10] or ~650°C [39].
Theoretical calculations predict quite well the upper temperature limit of the σ-phase for this system [39,40]; however, they fail to predict that the σ-phase loop closes at lower T.
Regarding the concentration range, the following values can be found in the literature (in at% V): 31.5 – 61 [21], 30-60 [42], 36-60 [3], 37–57 [32, 43], ~36 - ~56 [44], 40.5 – 59.1 [10], 39-54.4 [45]. In some cases only the lower range viz. 28.4 [46] or the upper one viz. 60 [47] are given.

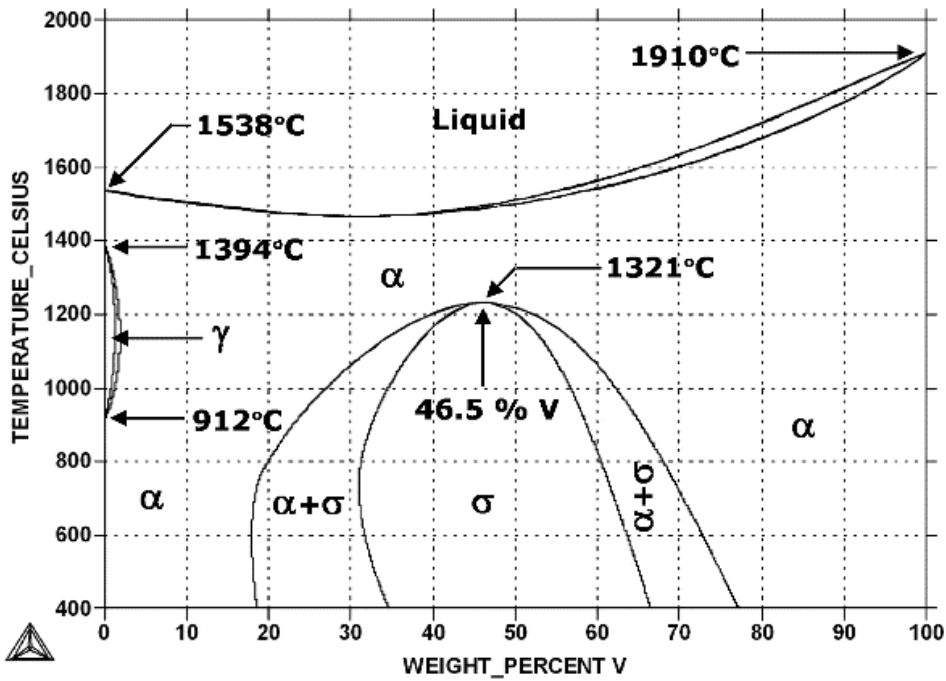

**Fig. 2.2**
*Phase diagram of a bulk Fe-V alloy system [36].*

## 3. IDENTIFICATION OF σ AND ITS STRUCTURE DETERMINATION

A first successful identification of the σ-phase goes back to early twenties of the previous century when Bain and Griffits found in a Fe-Cr-Ni alloy a hard, very brittle and non-magnetic phase which they termed the "B constituent" [10]. In other words, the hardness, which is usually by a factor of 3-4 higher than that of the α-phase from which it precipitates, can be taken, at least qualitatively, as a appropriate feature for identification of σ. However, for its quantitative identification and for determination of its crystallographic structure, X-ray (XRD) and neutron diffraction (ND) techniques are the most appropriate, though, in the case of σ in Fe-Cr and Fe-V systems, MS can also be regarded as the adequate method. Characteristic XRD patterns and MS spectra of α and σ phases recorded at room temperature are displayed in Fig. 3.1 to illustrate the statement.

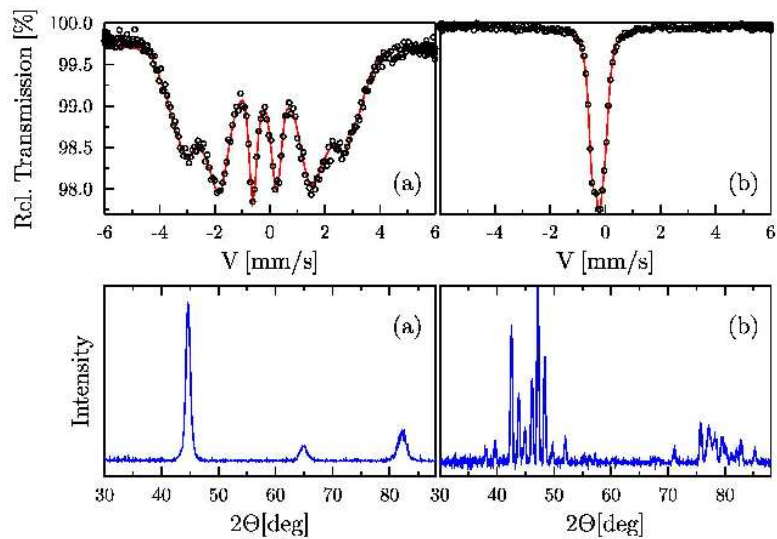

**Figure 3.1**
*$^{57}$Fe Mössbauer spectra recorded at room temperature on a quasi-equi-atomic Fe-Cr alloy for (a) α and (b) σ phases, and XRD patterns measured in the same conditions for (c) α and (d) σ phases.*

In the case of the Mössbauer spectra there is not only an obvious difference in their shape, which follows from the fact that at room temperature α is magnetic and σ is not, but there is also a huge difference in the isomer shift, *IS*, between the two phases, namely –0.11 mm/s [48], which, in the light of a scaling factor given in [49], corresponds to ~0.05 s-like electron, an unusually high value for metallic systems.

However, XRD and ND techniques have the advantage that they can also be used for determining structural properties: Concerning σ, its structure – tetragonal unit cell with 30 atoms, space group $D^{14}_{4}$ h –P42/mnm – was for the first time identified by Bergman and Shoemaker [13]. It can be also described as a D8$_b$ type compound (Strukturbericht).

A volume of the unit cell – see Fig. 3.2 - of the σ-Fe$_{535}$Cr$_{465}$ alloy, V$_o$ = 351.81 Å$^3$ is by ~5.5% less than that for the σ-Fe$_{50}$V$_{50}$ alloy [3].

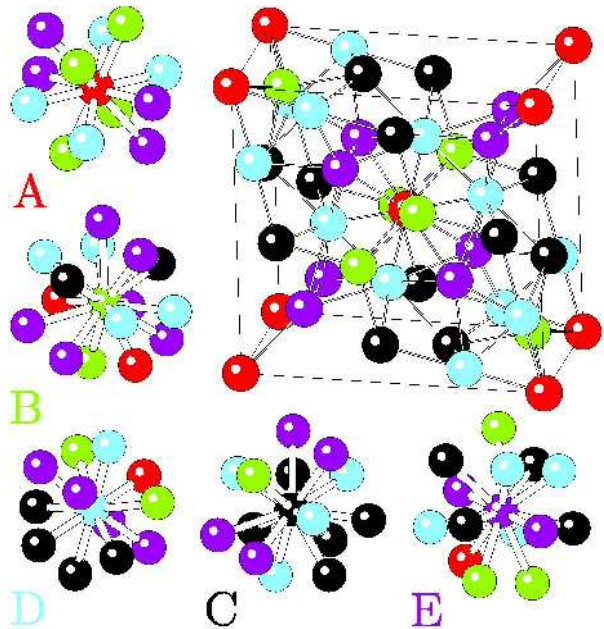

**Fig. 3. 2**
*The unit cell of σ, and five sub-lattices A, B, C, D and E with their nearest-neighbour shell atoms.*

The atoms are distributed over five sites A (I), B(II), C(III), D(IV) and E(V) with the occupations number (ON) equal to 2, 4, 8, 8 and 8, respectively. The sites have the following coordination numbers (CN): 12, 15, 14, 12 and 14. They are not equivalent not only because their CNs are different, but also because the nearest-neighbour (NN)-distances within the given sites are different. According to Bergman and Shoemaker [13] the average NN-distances for the sites A, B, C, D and E are as follows: 2.508, 2.701, 2.652, 2.526, 2.638, all figures are in Å. Consequently, the volume available around each site, V$_i$, is different. It varies in the order: V$_A$≈ V$_D$ < V$_C$ ≈ V$_E$ < V$_B$ i.e. it is correlated with CN. Also the symmetry of the sites is different: low for B and C sites, high for A and D sites and the highest for the E site. Therefore, the sites are not equivalent not only with respect to their crystallographic, but also with respect to their physical properties. The latter is especially crucial when using microscopic methods like MS or Nuclear Magnetic Resonance (NMR) as research tools.

Although Bergman and Shoemaker had succeeded in determining the unit cell parameters of σ, they had failed to determine the actual distribution of Fe/Cr atoms over the sites.

It was Yakel who had found that all five sites are "mixed" i.e. populated by both kinds of atoms constituting σ [18].

Similarly unclear situation with respect to site occupancy for σ was valid for the Fe-V system. According to ND measurements carried out on the $Fe_{40}V_{60}$ alloy [47], B sites were exclusively occupied by V atoms while all other four sites were mixed i.e. occupied by both Fe and V atoms. However, the actual occupation fraction depends on the sites and for Fe it is equal to: A = 85%, C = 17.5%, D = 65 % and E = 25%.

Recent ND study carried out on a series of σ-phase Fe-Cr and Fe-V alloys [19] has revealed that for both systems all sites are disordered, and the actual occupation numbers depend on the alloy composition as shown in Fig. 3.3. For the $Fe_{40}V_{60}$ alloy, in particular, the percentage occupation numbers for Fe atoms are as follows: A = 86%, B = 5%, C = E = 17.5% and D = 92%. In other words, a good agreement with the results reported in [47] exists only for the sites A and C, while the worst one for the site D. Eventually, the difference may be partly due to a different way of samples preparation, and partly to a poor purity of the sample studied by Kasper and Waterstrat, which was contaminated with 1.5 wt.% Si [47]. One should notice that the actual site occupation quite sensitively depends on the alloy composition. The sites C, E and B exhibit especially strong dependence. For example, the occupation number of the latter increases from 5% for x = 40 to 25% for x = 60.

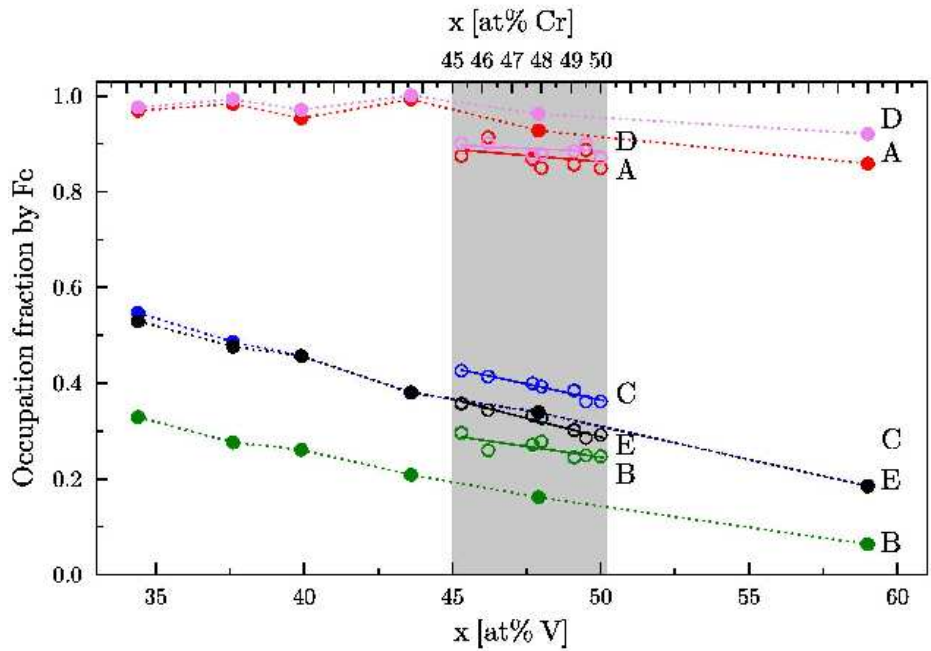

**Figure 3.3**

*Percentage site occupancy by Fe atoms in σ-phase Fe-Cr (open symbols) and in Fe-V (full symbols) alloy systems versus Fe concentration, x. The lines for Fe-Cr stand for the best fits to the data, while those for Fe-V to guide the eyes. The plot has been made based on the data reported elsewhere [19] as well as the non-published data measured by the authors.*

From the data presented in Fig. 3.3 it is evident, that there are both similarities as well as differences between the two systems as far as the site occupancy is concerned. Regarding the former, Fe atoms predominantly occupied the sites A and D in both systems, though with a systematically higher probability for Fe-V. For both systems the probability of finding Fe atoms on the site B is the lowest, yet for Fe-V significantly lower. Regarding the differences, the probability for finding Fe atoms on sites C and E is the same for Fe-V, while it is distinct for Fe-Cr.

Our recent ND experiments have also enabled determination of the lattice parameters, $a$ and $c$, hence the volume of the unit cell, $V_\sigma$. The corresponding results are presented in Fig. 3.4.

It is clear that the unit cell of σ-FeV is considerably larger than the one of σ-FeCr. The difference increases with the increase of V/Cr content, $x$. Contrary to the behaviour of $a$ and $V$, $c$ decreases with $x$ for Fe-Cr.

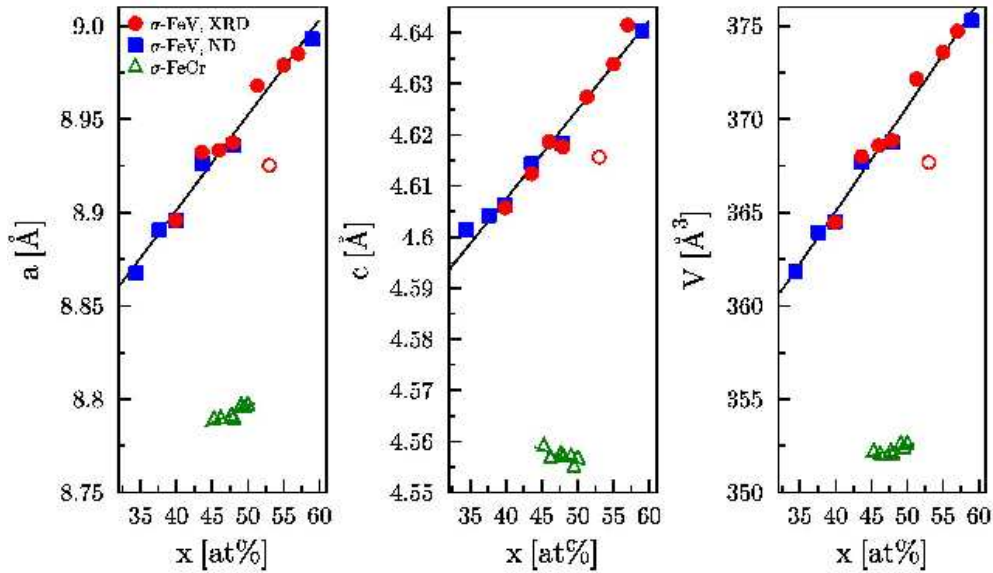

**Fig. 3.4**
*Lattice parameters, a and c, as well as the unit cell volume, V, for σ-phase Fe-Cr and Fe-V alloys versus Cr/V concentration, x [redrawn after Ref. 75].*

## 4. KINETICS OF α-to-σ TRANSFORMATION

Kinetics of the α-to-σ transformation can be studied with any method that is able to distinguish between α and σ phases. There are two main reasons for such studies: (1) scientific, aimed at revealing the mechanism responsible for the transformation, and (2) practical, aimed at determining conditions that eliminate or, at least, retard the formation of σ. The latter is related with a technological importance of materials like SS, which drastically deteriorate under precipitation of σ, hence loose their useful properties such as resistance to a high-temperature corrosion and toughness. As SS are usually composed of several elements, Fe-Cr alloy being their main ingredient, in this kind of kinetic investigations one studies the effect of other elements on the kinetics.

The first investigation of the kinetics of the σ-phase formation was carried out by Baerlecken and Fabritius on a Fe-Cr sample containing 48.3 wt% Cr, 0.93% Si, 0.18% Mn and 0.16% C [25]. The α-to-σ transformation was investigated with a magnetization method. A relative amount of the precipitated σ-phase was determined from the saturation magnetization recorded at RT. The kinetics was investigated in the temperature range of 570 to 860°C, where 100% sigma could have been obtained. At T > 860°C it was not possible to get 100% of sigma. Different virgin samples of the same ingot exhibited different transformation kinetics, most likely due to a lack of a chemical homogeneity in the ingot (its mass was ≈8 kg).

Samples having different size of grains, <d>, had similar kinetics. The difference between that for <d> = 210 μm was only slightly slower than that for <d> = 125 μm, but the reason for that obviously lies in fact that the ratio between the surface and bulk atoms is not large enough for grains of this size to show the difference in the kinetics. The latter shows up for nano-crystalline samples – see the bottom part of this section.

According to the above-mentioned study, the α-to-σ transformation needed some incubation time, $t_i$ that as the time of total transformation, $t_t$ depended on the annealing temperature, $T_a$. For example:

$T_a$ =570 C, $t_i$ = 450 h, $t_t$ = 2200 h
$T_a$ = 780 C, $t_i$ = 1h, $t_t$ = 14h.

Temperature, at which the transformation rate was maximum, $T_m$ = 750° C.

The kinetics curves had S-shape, characteristic of nucleation and growth processes. For such processes, the kinetics was initially slow, then fast, and again slow in the final stages. For their description the Johnson-Mehl-Avrami-Kolmogorov (JMAK) equation was used [50]:

$$A_\sigma(t) = 1 - \exp[-(kt)^n] \qquad (4.1)$$

where $A_\sigma(t)$ is the amount of the σ-phase precipitated after time $t$ of annealing, $n$ is the Avrami exponent ($n$ = 4 according to Mehl, and $n$ =5/2 according to Zener) that gives information about a mechanism responsible for the transformation, and $k$ is a time constant which, via Arrhenius equation, can be related to the activation energy, $E_a$, through the following equation:

$$k = k_o \exp(-\frac{E_a}{RT}) \qquad (4.2)$$

Where $R$ is the gas constant, $T$ is temperature and $k_o$ a prefactor.

It was found in this study that the Avrami exponent had values depending on the annealing temperature as follows: $n$ = 4.6 for T ≤ 600° C; $n$ = 3.5 for T = 670° C, $n$ = 2.9 for T = 715 and 780° C, $n$ = 2.5 for T = 820° C and $n$ = 4.6 for 860° C.

The values of the activation energy determined for different temperature intervals and three stages of transformation are displayed in Table 4.1.

**Table 4. 1** Activation energy, $E_a$ [kJ/mol] [25]

| T [° C] | $E_a$ [kJ/mol] | | |
|---|---|---|---|
| | $t_i$ | $t_{1/2}$ | $t_t$ |
| 570-600 | 372 | 372 | 343 |
| 600-670 | 280 | 238 | 217 |
| 670-715 | 167 | 201 | 155 |

The authors of that study also found for the samples that have undergone the first cycle of transformation (α-to-σ-to-α-σ), $E_a$ = 188 kJ/mol, independently on the temperature and time.

In the authors' opinion, the smaller value of $E_a$ is due to that these samples need less energy to create seeds of the σ-phase in comparison with the virgin samples.

Concerning MS, it can be readily used for a quantitative study of the issue in two ways; the amount of the precipitated σ-phase, $A_\sigma(t)$, can be determined either (i) classically i.e. from a spectral area or (ii) newly viz. from the average isomer shift, *<IS>*, as illustrated in Fig. 4.1.

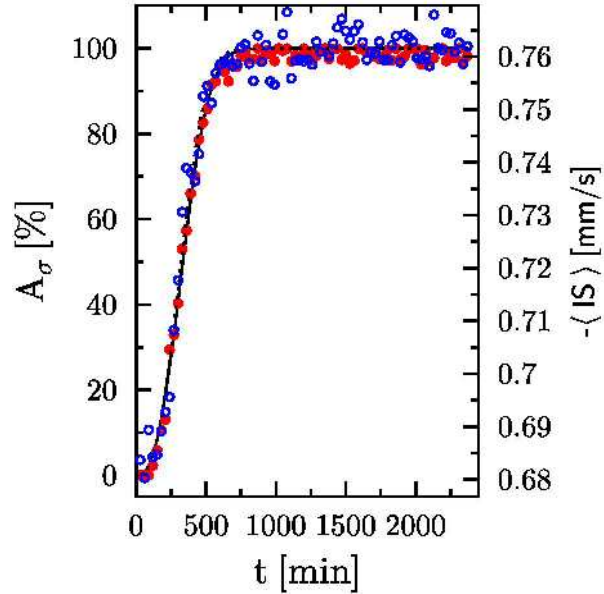

**Fig. 4.1**
*Typical dependence of the (left-hand scale) fraction of precipitated σ-phase, $A_\sigma$, and (right-hand scale) average isomer shift, <IS>, as a function of annealing time, $t_a$. The solid line stands for the best fit to the data in terms of eq. (1). The plot has been made based on the data published elsewhere [60].*

The first application of the approach (i) to investigate the issue was that done by Japa et al. [51], in which the α-to-σ transformation was only qualitatively investigated *ex situ* on a sample of $Fe_{53.8}Cr_{46.2}$ alloy annealed at three different temperatures. The study revealed that the fastest transformation rate took place at 700°C. Since then several papers were devoted to the subject. Kuwano determined for the first time the value of the activation energy for the formation of σ in Fe-Cr using MS, $E_a$ = 193 kJ/mol [52]. Costa and Dubiel studied the influence of Sn on the formation of σ in quasi-equiatomic Fe-Cr alloys [53]. They found that an increase of Sn content retards the transformation process, partly because Sn atoms precipitate on grain boundaries blocking thereby the preferable sites for the formation of σ. In turn, Waanders and co-workers using the traditional method studied the effect of Mo, and they found that the increase of Mo content from 0 to 5.5 at% increases the activation energy from 200 kJ/mol for $Fe_{50.7}Cr_{49.3}$ sample to 283 kJ/mol for $Fe_{52.8}Cr_{40.5}Mo_{5.5}$ alloy [54,55]. The effect of Ti on the formation of σ at $T_a$ = 680-700°C was investigated with MS for $Fe_{53.8}Cr_{46.2}$ alloys with different size of grains [56, 57]. $E_a$ = 207 kJ/mol was determined for a small grain $Fe_{53.8}Cr_{46.2}$ alloy. Addition of Ti decreases $E_a$ up to a certain critical content equal to ~1 at% Ti for coarse-grained samples and ~1.2 % Ti for fine-grained ones. For greater concentrations the effect is to slow down the transformation. In terms of $E_a$ the most effective acceleration of the transformation takes place for ~0.3 at% Ti, where Ea is smaller by ~ 19 kJ/mol, and the most effective retardation occurs for 3 at% Ti where $E_a$ is enhanced by ~10 kJ/mol. Blachowski and co-authors determined the activation energy based on method (i) using various approaches [58]: (a) the rate constant, (b) the length of time between two different stages, (c) the time to a given fraction, (d) the maximum transformation rate and (e) the two-temperature kinetics. They showed that the actual value of $E_a$ depends on the method used to determine it. In particular, for the $Fe_{53.8}Cr_{46.2}$ alloy, $E_a$ ranges between 197 and 260 kJ/mol, with the average value of 196 kJ/mol, and between 119 and 183 kJ/mol for the $Fe_{53.8}Cr_{46.2} -$ 0.5%Ti alloy, with the average value of 153 kJ/mol. Whatever method used, the activation energy for Ti-doped alloys was smaller, which was consistent with the observation that the addition of 0.5 at% Ti accelerated the transformation. It should be mentioned here that the Mössbauer spectroscopic study gave evidence that in early stages of transformation in Fe-Cr-Sn alloys the isomer shift of the precipitated σ-phase was not constant but was changing continuously with the amount of precipitated σ, reaching a constant value characteristic of the pure σ when its amount was greater than ~30% [59]. The reason for such behaviour is not clear but it may reflect the effect of size of grains of σ or some re-arrangement of atoms within the new crystallographic structure in the course of its growth. The occurrence of the ordered superstructure (B2) is also possible, as it is known to happen in the Fe-V system [46]. Alternatively, as we have shown for the first time [56, 60], the process of α-to-σ transformation can be equally well described in terms of an average isomer shift, <IS>. Figure 4.1 gives evidence that both approaches are equivalent as far as the quantitative description of the kinetics analysis is concerned. However, the advantage of using the approach (ii) lies in that <IS> can be determined much easier and more accurately than the relative area, $A_\sigma$. Finally, concerning the study of the kinetics in bulk alloys by means of MS, its *in situ* application [56] should be mentioned as a possible alternative to the conventional *ex situ* mode. In that study we have also determined a ratio between the Lamb-Mössbauer factor (recoil-free fraction), $f$, for σ and α as equal to 1.15. The result means that the corresponding correction is needed when determining the kinetics from the spectral area. In other words, the relative amount of the σ-phase, $A_\sigma$, after the above-mentioned correction can be determined from the following formula:

$$A_\sigma = \frac{S_\sigma f_\alpha}{S_\sigma f_\alpha + S_\alpha f_\sigma} \tag{4.3}$$

Where $S_i$ stands for the spectral area associated with the phase σ (i = σ) or α (i = α). The recoil-free fraction is denoted by $f_i$ and its subscribe has the same meaning as that in $S$.

Last but not least, the applications of MS in the study of the kinetics of α-to-σ phase transformation in mechanically alloyed samples are worth mentioning [61]. Using the traditional *ex situ* transmission mode Costa and co-workers investigated a series of Fe-Cr-Sn alloys with Sn content as high as ~6 at%, the latter achieved thanks to the high-energy milling conditions. The results obtained in that study permitted to conclude that the precipitation rate of σ in the mechanically alloyed samples is higher than the one in the alloys of similar compositions but prepared by a conventional melting process. In the authors' opinion this follows from the higher fraction of grain boundaries per sample volume in alloys with nanometre-sized grains, hence prepared by a mechanical attrition.

## 5. KINETICS OF σ-to-α TRANSFORMATION

According to the phase diagram, the reverse transformation i.e. from σ to α can be obtained by an isothermal annealing of σ-phase samples at temperatures higher than the critical one, which in the case of Fe-Cr system is equal to ~820°C [24]. Such study is both of fundamental as well as of technological interest. The former is important from the view-point of our knowledge of phase-transition phenomena and underlying mechanisms in alloy systems with a complex crystallographic structure like that of σ, while the latter may have, at least potentially, practical applications, namely for a recovery of objects whose properties have been deteriorated by the presence of σ. To our best knowledge, Baerlecken and Fabritius carried out such studies for the first time [25]. The authors revealed that this process is different than α-to-σ one in that there is no incubation time; instead the transformation starts instantly upon annealing.

Concerning the application of MS to study the issue, the first one was devoted to the investigation of such transformation in Fe-Cr alloys induced by ball milling [62, 63]. Both nano-grained as well as policrystalline samples of σ-$Fe_{54}Cr_{46}$ alloys were ball-milled in argon atmosphere. In both cases the milling resulted in a transformation into α, the rate of transformation being faster for the nano-crystaline samples. However, the formation of a ferromagnetic bcc-phase was accompanied by another phase, which was identified as amorphous. The kinetics of this process, which is described in detail elsewhere [64], can be quantitatively described also in terms of the JMAK equation with similar values of the time constant, *k*, but with different values of the Avrami exponent *n* in comparison with those for the reversed process i.e. transformation of α into σ.

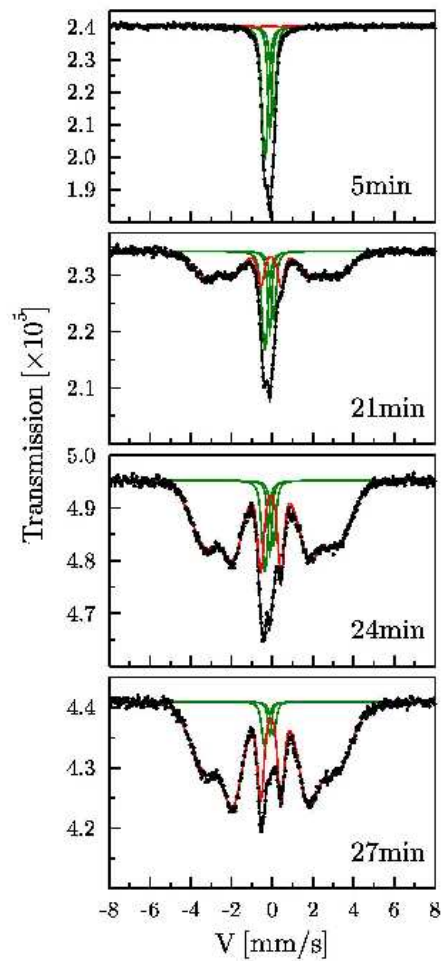

**F**ig. 5.1
*$^{57}$Fe Mössbauer spectra recorded on the σ-$Fe_{51}Cr_{49}$ sample annealed at 845$^{O}$C for different periods shown. Subspectra corresponding to α and σ are indicated.*

First qualitative study of this reverse transformation using MS was recently carried out in the *ex situ* mode on two samples of σ-FeCr alloys viz. $Fe_{51}Cr_{49}$ and $Fe_{53.8}Cr_{46.2}$ [65]. The transformation process was promoted by the isothermal annealing at various temperatures higher than the critical one (821°C), namely within the range of 825 and 860° C. Examples of the spectra recorded on the sample $Fe_{51}Cr_{49}$ annealed at 845°C for different periods are displayed in Fig. 5.1. The spectra were analyzed to get a fraction of the precipitated α-phase, $A_\alpha$, using the procedure outlined in detail elsewhere [65]. The time dependence of the precipitate fraction of the α-phase, $A_\alpha$, shown in Fig. 5.2, could have been successfully described in terms of the JAMK equation. From the latter kinetics parameters $n$ and $k$ were evaluated. The activation energy of $E_a$ = 1013 ±90 kJ/mole was calculated via Arrhenius law – eq.(4.2). As this value is by a factor of 5 higher than the activation energy found for the reversed process [52], it follows that the two processes are characterized by a strong thermodynamic asymmetry. The study has also clearly demonstrated that the Avrami exponent $n$ depends on temperature changing its value from ~1.5 for T = 825°C, indicating a heterogeneous process, to ~4 for T > ~840°C indicating a homogenous one.

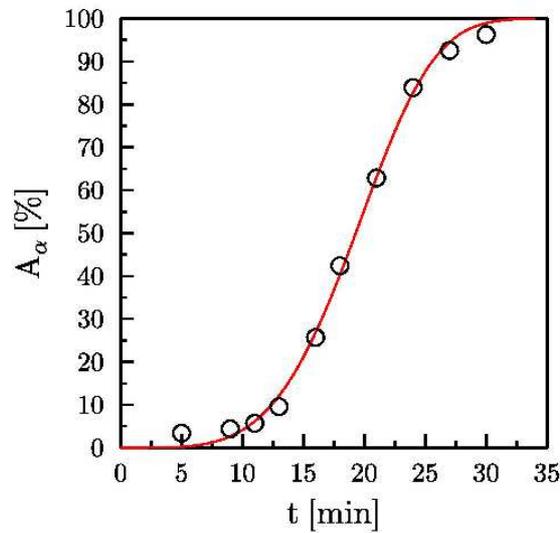

**Fig. 5. 2**
*Kinetics of the σ-to-α transformation for the sample of σ-$Fe_{51}Cr_{49}$ annealed at 845°C. The solid line represents the best-fit to the data obtained in terms of the JMA equation.*

In that study we also found some evidence from the ratio of the second and third lines, $C_2/C_3$, in the spectra that the orientation of the easy magnetization axis depends strongly on the amount of the α-phase precipitated in the sample, $A_\alpha$. Namely, it increases from ~1 to ~3 which means that the preferred orientation of the magnetization continuously changes with temperature from the "out-of-plane" (the magnetization axis perpendicular to the sample's surface) towards the "in-plane" (the magnetization axis parallel to the sample's surface) position as the transformation process proceeds.

## 6. DEBYE TEMPERATURE

The most popular model for describing a lattice dynamics is that due to Debye, according to which the phonon density of states, *PDOS,* is a parabolic function of frequency, $\omega$, i.e. *PDOS* $(\omega) = C \cdot \omega^2$, for $\omega \leq \omega_D$, and *PDOS* $(\omega) = 0$, for $\omega > \omega_D$ (C being a constant). In other words, the oscillating frequency of lattice atoms ranges from zero to a maximum value of $\omega_D$, called Debye frequency, which defines the so-called Debye temperature through the relation $\hbar \cdot \omega_D = k_B \cdot \Theta_D$. This characteristic temperature is the most frequently given scalar characteristics of the lattice dynamics. There are numerous methods applied to determine $\Theta_D$. It should be, however, remembered that the actual value of $\Theta_D$ depends on a particular method used to determine it, which reflects the fact that the Debye temperature is not a physical quantity, but rather a parameter introduced to describe the lattice dynamics via the Debye model. The sensitivity of $\Theta_D$ to the method applied to evaluate it follows from the fact that *PDOS* in a real system has much more complex shape than the parabola – see Fig. 6.1, and various experimental methods probe different parts of *PDOS* resulting in different $\Theta_D$-values. To illustrate this issue numerically let us consider the cases of a pure Fe and pure Cr. Concerning the former $\Theta_D$ ranges between 477 K as determined from the elastic constant [67], and 418 K as estimated from the X-ray diffraction experiment [68]. Regarding the latter, the amplitude of the difference in $\Theta_D$ is even larger, because it spans between 630 K, as known from the specific heat measurements [69], and 466 K, as found from the Young's modulus [70].

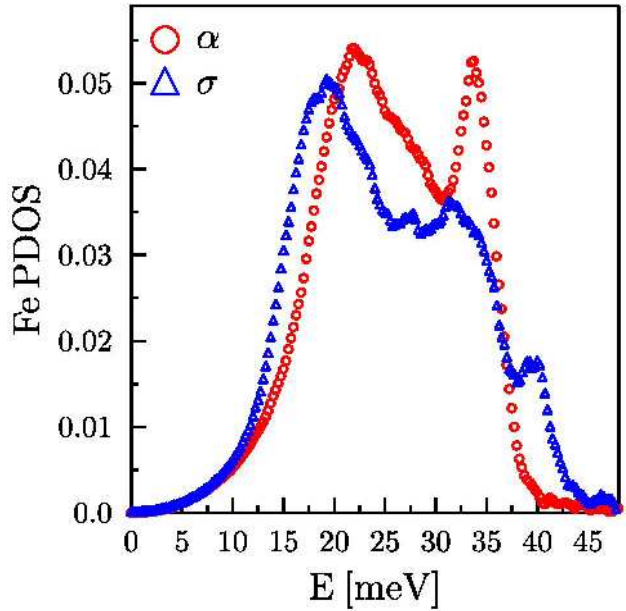

**Fig. 6.1**
*Phonon DOS as measured on $^{57}$Fe atoms at 298 K for the $\alpha$- (circles) and the $\sigma$-phase (triangles) on the $Fe_{52.5}Cr_{47.5}$ samples [redrawn after Ref. 66].*

Consequently, a comparison of $\Theta_D$-values obtained for a given sample with different approaches seems not to be reasonable. Rather a systematic study of a given system with one particular approach makes much more sense, and an eventual comparison between the results obtained with various methods should be done on the basis of normalized $\Theta_D$-values. The Mössbauer spectroscopy plays a special role among various experimental methods that can be used to determine $\Theta_D$ as it appears to be the only one allowing determination of $\Theta_D$ from one series of measurements but based on two different physical quantities viz. the centre shift, *CS*, and the recoil-free fraction, *f*. The former can, within the Debye model, be expressed as follows:

$$CS(T) = IS(0) - \frac{3k_B T}{2mc}\left[\frac{3\Theta_D}{8T} + \left(\frac{T}{\Theta_D}\right)^3 \int_0^{\Theta_D/T} \frac{x^3}{e^x - 1} dx\right] \tag{6.1}$$

Where *M* is the mass of the $^{57}$Fe nucleus, $k_B$ is the Boltzmann constant, *c* is the velocity of light, and *IS (0)* is the isomer shift (charge-density), which to the first-order approximation can be treated as temperature independent.

On the other hand, the corresponding formula for *f* reads as:

$$f = \exp\left\{-\frac{6E_R}{k_B \Theta_D}\left[\frac{1}{4} + \left(\frac{T}{\Theta_D}\right)^2 \int_0^{\Theta_D/T} \frac{x}{e^x - 1} dx\right]\right\} \quad (6.2)$$

Where $E_R$ is the recoil-free energy. Values of $\Theta_D$ obtained with the two approaches are usually different, which follows from the fact that *CS* is related to the mean-squared velocity of the vibrating atoms, while *f* to their mean-squared displacement from the equilibrium position. For example, in the case of iron, $\Theta_D$ = 421K as determined from *CS*, and $\Theta_D$= 358K as found from *f* [71]. It should be realized here that from an experimental viewpoint, determination of *f* is much more demanding than finding *CS*, hence the former is less frequently used in practice. For the above-mentioned reason, we applied the easier and more accurate approach for determining $\Theta_D$ in our study of σ-phase Fe-Cr and Fe-V alloys [72-75]. For each composition, a series of 15-16 spectra was recorded in the temperature interval of 60-300 K. Some examples of them can be seen in Fig. 6.2. In the case of complex spectra i.e. composed of several subspectra, like in the present instance, one determines an average value of *CS*, *<CS>*. Details of the procedure used for determining *<CS>* for the σ-phase are given elsewhere [72].

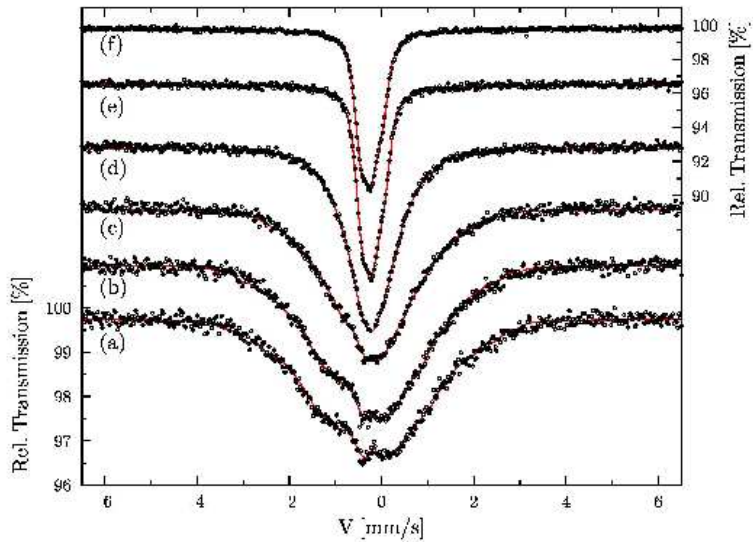

**Fig. 6.2**
*[57]Fe Mössbauer spectra recorded on the σ-Fe$_{60}$V$_{40}$ sample at (a) 4.2 K, (b) 50 K, (c) 100 K, (d) 150 K, (e) 200 K, and (f) 250 K. The full lines stand for the best-fit spectra. The left-hand scale refers to the spectrum (a) and the right-hand one to the spectrum (f) [75].*

A typical temperature dependence of *<CS>*, is illustrated in Fig. 6.3, while in Fig. 6.4 the values of $\Theta_D$ derived from such procedure are displayed.

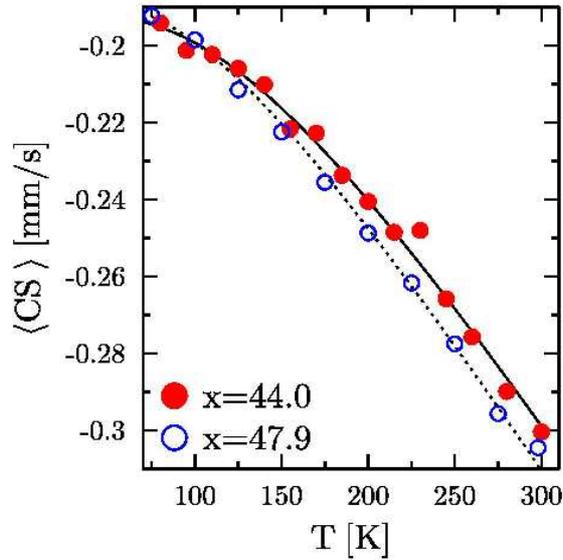

**Fig. 6.3**
*Examples of the temperature dependence of the average center shift, <CS>, for two σ-$Fe_{100-x}V_x$ samples with the x – values shown. The best fits to the experimental data in terms of eq. (6.1) are indicated by a full and a dotted line, respectively [75]*

Concerning the Fe-Cr system, where σ exists in a much narrower range than that in the Fe-V system, $\Theta_D$ exhibits a linear and rather steep increase with x viz. from ~415 K at $x \approx 44$ to ~480 K at $x \approx 49$. These figures can be compared with the value of 398 K derived from $f$ as determined from the room temperature *PDOS* recorded on $Fe_{52.5}Cr_{47.5}$ alloy [66]. It is interesting to note that for the α-FeCr alloys of similar compositions, $\Theta_D$ parallels the behaviour found for σ, though its values are shifted downwards by ~10 K [76].

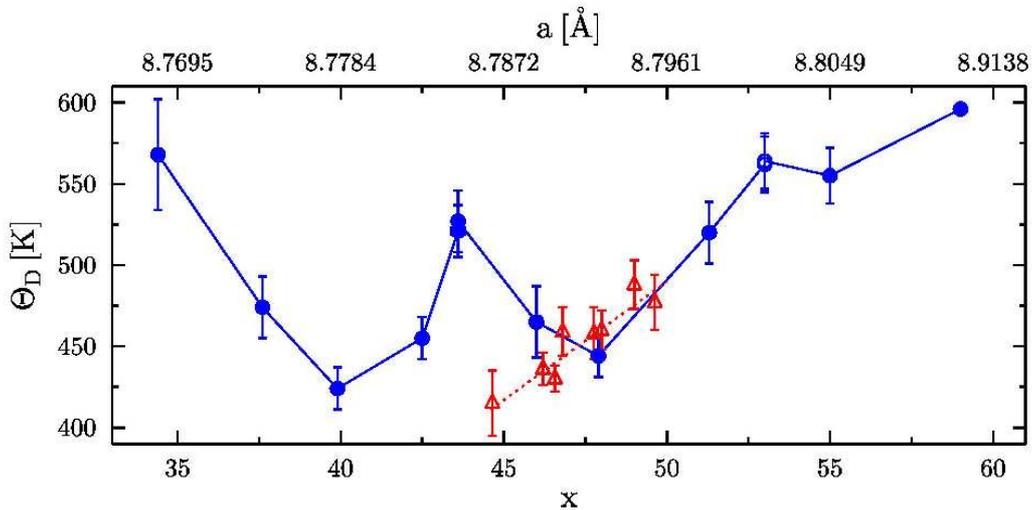

**Fig. 6.4**
*Dependence of the Debye temperature, $\theta_D$, on vanadium (circles) and chromium (triangles) concentration, x, and on the lattice constant, a (full circles for the bulk samples and open circle for the nanocrystalline one). Solid and dotted lines are only guides to the eyes. The error bar in x has a size of the symbol [75].*

In the Fe-V system, where the range of σ occurrence is much wider, $\Theta_D$ shows a non-monotonic behaviour as a function of composition with its extreme values between ~425 K for $x \approx 40$ and ~600 K for $x \approx 60$. A local maximum of ~525 K was found to exist at $x \approx 43$.

# 7. CURIE TEMPERATURE

MS can be readily used to determine a characteristic temperature at which a magnetic order sets in. This can be either the Curie temperature, $T_C$, for ferro- or ferrimagnetic materials or the Néel temperature, $T_N$, for antiferromagnetically ordered ones. The ability of MS for such applications follows from that the magnetic hyperfine field, $B$, which is related with the magnetization of the material, can characterize the magnetic materials. However, one has to remember first, that $B$ is defined locally i.e. within the nucleus of the probe atoms, and the resulting splitting of the spectrum via the nuclear Zeeman effect is not sensitive to the orientation of the magnetic moment associated with a given atomic site. Consequently, one observes $B$ in antiferromagnetic materials, despite their net magnetization is zero. Second, $B_{hf}$ is not strictly proportional to the magnetic moment, $\mu$ [77], and also the time scale involved in the measurements of $B$ and $\mu$ differs by several orders of magnitude. Consequently, if we limit ourselves only to the ferromagnetic materials, what is the case of σ in Fe-Cr and Fe-V alloys, we have to realize that $T_C$ – s derived from MS and those obtained from the magnetization measurements are not exactly the same quantities. As a rule, $T_C$ determined from MS is higher.

Methodologically, there are two ways of determining $T_C$ using MS viz. from the temperature dependence of: (i) the width of the spectrum, $G(T)$, and (ii) $B(T)$. The former approach is suitable for magnetically weak materials i. e. such for which the spectrum has no well-resolved structure. Otherwise, the latter is the right method. Below, both approaches will be presented in more detail.

The magnetism of σ-FeCr alloys can be regarded as very weak, because the maximum magnetic moment per Fe atom, $\mu_{Fe}$, as determined from the magnetization measurements is equal to ~0.3 $\mu_B$ [7]. Correspondingly small is $B$, and, consequently, the Mössbauer spectrum exhibits merely a broadening with respect to its width in a paramagnetic state – see Fig. 7.1.

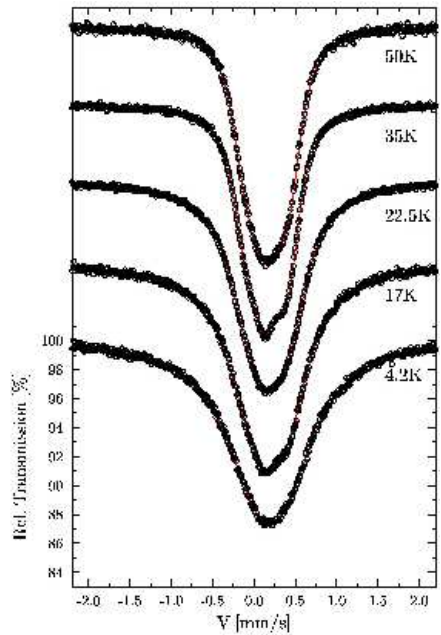

Fig. 7.1
$^{57}$Fe Mössbauer spectra recorded on σ-$Fe_{53.8}Cr_{46.2}$ at various temperatures between (a) 4.2 K and (d) 300 K.

In such a case, the spectrum can be fitted as composed of one subspectrum, whose line width, *G*, is treated as a free parameter. By plotting *G* versus *T*, as illustrated in Fig. 7.2 for the Fe-Cr case, one can find $T_C$. This procedure can be also applied to determine $T_C$ in σ-FeV samples, which are also magnetically weak.

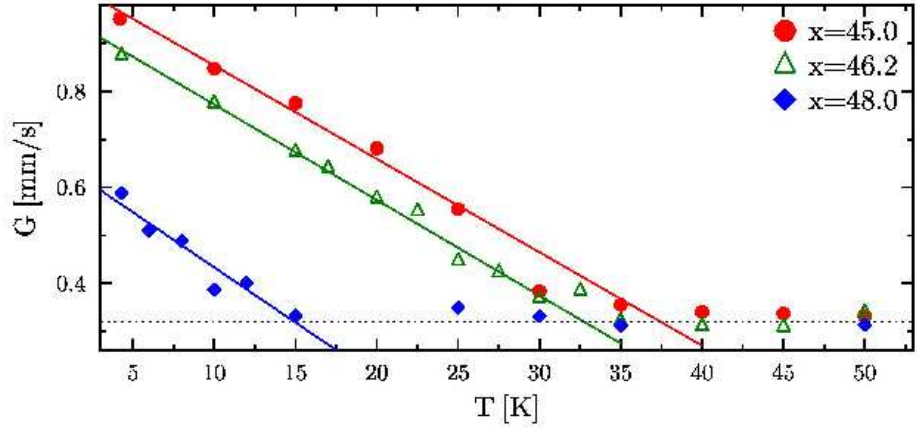

**Fig. 7.2**
*The full line width at half maximum, G, versus temperature for three different samples of σ-FeCr$_x$ with (a) x =45.0, (b) x = 46.2 and (c) x = 48.0. The intersection with the horizontal line (dotted) defines the Curie temperature, $T_C$.*

However, in the Fe-V alloy system, as already mentioned, σ can be obtained in a much wider range of composition than that in Fe-Cr. For σ-FeV samples with lower content of vanadium, their magnetism is strong enough to cause an appreciable broadening in the spectra – see Fig. 7.3. In particular, a sample with vanadium concentration equal to 34.4 at% is magnetic even at room temperature, and $\mu_{Fe} \approx 0.9$ $\mu_B$ [8]. In this case $T_C$ can be easily determined from *B(T)*. Yet, as it clearly follows from Fig. 7.3, the structure of the spectra does not allow to analyze them in terms of components (subspectra) corresponding to different sites. Instead, one can model them using a hyperfine field distribution (HFD) method to get the distribution of the hyperfine field, *P(B)*, where *P(B)dB* is a probability of finding the hyperfine field *B* in the interval of *[B,B+db]* [78]. In our analysis of the σ-FeV spectra in order to account for their visible asymmetry, it was assumed that the hyperfine field was linearly correlated both with the isomer shift and the quadrupole splitting. By integration of the *P(H)*-curves, the average value of B, <B>, was obtained. An example of a <B>(T) – dependence, from which $T_C$ could be determined, is shown in Fig. 7.4. The value of $T_C$ = 323 K as found for the σ-Fe$_{65.6}$V$_{34.4}$ is record-high for any magnetic σ-phase known to date. The corresponding value derived from the magnetization measurements is 315 K [8].

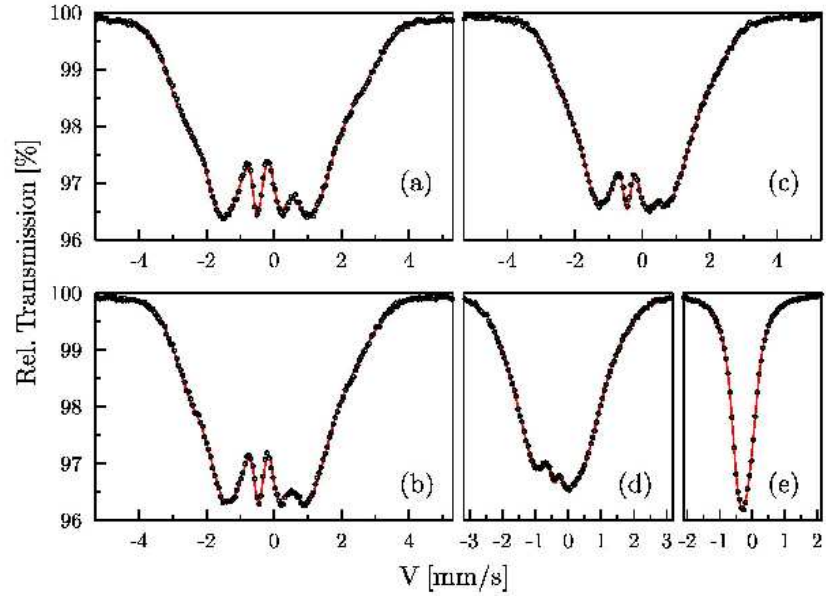

**Fig. 7. 3**
*$^{57}$Fe Mössbauer spectra recorded on σ-Fe$_{65.4}$V$_{34.4}$ at various temperatures between (a) 4.2 K and (d) 330 K. The solid lines represent the best fit to the spectra [redrawn from Ref. 8].*

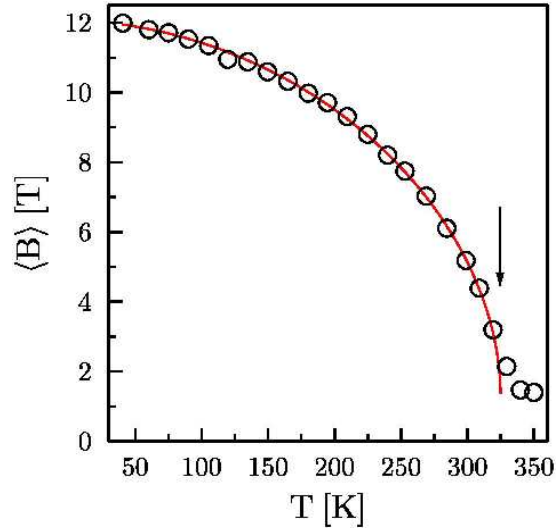

**Fig. 7. 4**
*Temperature dependence of the average hyperfine field, <B>. as derived from the spectra measured on σ-Fe$_{65.4}$V$_{34.4}$. The arrow shows the Curie temperature, $T_C$ . The plot has been made based on the data published elsewhere [8].*

## 8. HYPERFINE FIELD AND MAGNETIC MOMENT

A knowledge of such relationship is a prerequisite for carrying out the correct rescaling of *B* into underlying magnetic moment, $\mu_{Fe}$, and vice versa. The simplest possible connection between the two quantities is a linear one i.e. $B = A \cdot \mu_{Fe}$, where *A* is known as the hyperfine coupling constant. Despite several theoretical calculations showed that the linear relation does not hold, in practice many investigators applied it. Furthermore, they used ~15 $T/\mu_B$ for $A$ (the value derived from the pure metallic Fe) – see [77] and references therein. In our recent paper we gave experimental evidence, for several alloys and compounds of iron, that there is no one universal value of $A$ [77]. Here we limit ourselves to the case of the relationship for σ in Fe-Cr and Fe-V systems. The relationship was obtained based on our own measurements of the magnetization and those of the hyperfine fields. Figure 8.1 shows the results.

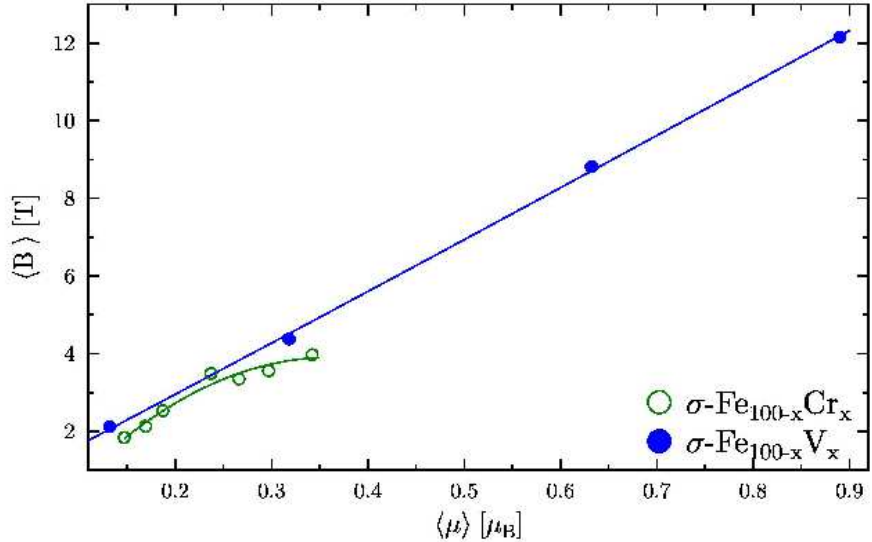

**Fig. 8.1**
<$B_{hf}$> - $\mu_{Fe}$ relation as deduced from magnetization and hyperfine field measurements for σ in Fe-Cr and Fe-V alloy systems . The plot has been made based on the results published elsewhere [8].

Clearly, the relation is linear for Fe-V, with $A$= 14.3 $T/\mu_B$, hence almost the same as for the metallic iron, but there is some evident departure from the linearity for Fe-Cr. In the latter case $A$ lies between ~8 $T/\mu_B$ and ~16 $T/\mu_B$. It should be mentioned first, that the relation found for σ-FeV alloys is exceptional [77], and second, that for both systems the values of $\mu_{Fe}$ have been calculated assuming only Fe atoms contribute to the magnetization. One has to be aware that if this assumption is not correct, then the relation shown in Fig. 8.1 is not correct either.

## 9. THEORETICAL CALCULATIONS

Theoretical calculations for σ-phases are rather seldom which follows from the fact that such calculations, although challenging, are very demanding due to the complex crystallographic structure of σ and a lack of stoichiometry. Yet fewer are those devoted to a determination of physical properties of this phase. Among the latter we are aware of only four papers dedicated to σ in Fe-Cr and Fe-V systems [79-82], but only ours are relevant to the subject of this contribution, so the results discussed in these papers will be mentioned here.
Thus in [79] the electronic structure of a σ-FeCr compound in a paramagnetic state was calculated for the first time in terms of the isomer shifts and quadrupole splittings. The former were calculated using the charge self-consistent Korringa–Kohn–Rostoker (KKR) Green's function technique, while the latter were estimated from an extended point charge model. The calculated quantities combined with recently measured site occupancies [18] were successfully used, as illustrated in Fig. 9.1, to analyze a Mössbauer spectrum recorded at room temperature using only five fitting parameters, namely: background, total intensity, line width, *ISO* (necessary to adjust the refined spectrum to the used Mössbauer source) and the *QS* proportionality factor. Theoretically determined changes of the isomer shift for the σ-FeCr sample were found to be in line with the corresponding ones measured on a α-FeCr sample [48].

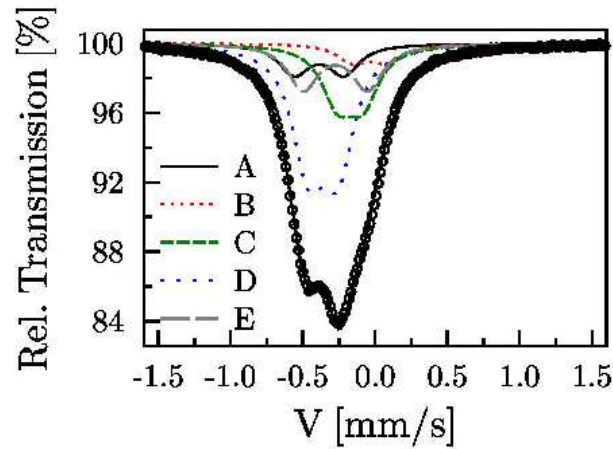

**Fig. 9. 1**
*$^{57}$Fe Mössbauer spectrum of the σ-$Fe_{53.8}Cr_{46.2}$ alloy recorded at 300 K (dots) compared to the calculated one (solid line). The subspectra related to the non-equivalent sites are indicated by dotted and dashed lines .*

As illustrated in Fig. 9.2, the calculations enabled determination of the Fe-site charge-density (isomer shift) for each sublattice as a function of the average number of Fe atoms, Fe-NN, within the nearest-neighbour shell. The results clearly show that the charge-density is characteristic of a given site, and for the each site, it fairly linearly decreases with Fe-NN, which may be interpreted as evidence for a charge flow away from Fe-atoms. Furthermore, the data displayed in Fig. 9.1 give clear evidence that the highest charge-density experience the sites A and D, while the lowest one the sites B. The extreme difference is roughly independent of Fe-NN and it is equivalent to ~0.3 mm/s or ~0.15 s-like electron, using the scaling factor outlined in [49].

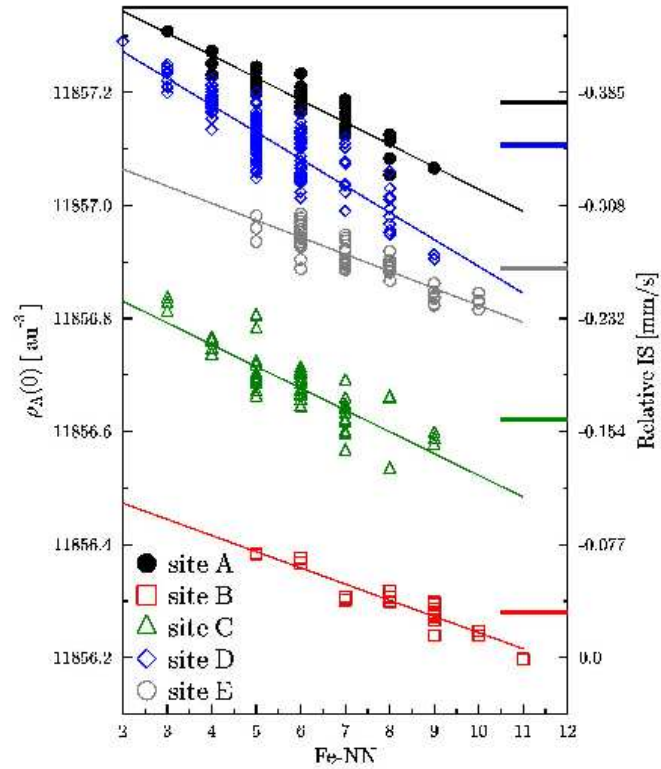

**Figure 9.2**
*Electron densities at the nucleus of Fe calculated for each lattice site of the FeCr σ -phase (left-hand scale) or, equivalently, isomer shifts relative to the source, IS (right-hand scale). Solid lines represent the best linear fits between the $\rho_A$ -values and the number of Fe-NN atoms. Different symbols denote the calculated $\rho_A$- values corresponding to nonequivalent sites. The average values of the charge-density (or isomer shifts relative to the source, $IS_{av}$) are indicated on the right-hand scale of the plot. The plot has been made based on the results published elsewhere [79].*

These results are in disaccord with those obtained by Gupta and co-authors [83] who analyzed their spectra in terms of only three subspectra assuming the sites B have the same spectral parameters as C, and sites A the same as D. Although, the statistical quality of such analysis was formally good, values of both spectral parameters and the relative population of the sites are not. Concerning the latter, a disaccord with the corresponding figures obtained by Yakel [18] reaches up to 11%. Also the values of the isomer shift, *IS,* and the quadrupole splitting, *QS*, as determined from that analysis significantly differ from our recent calculations reported in [79]. In particular, the greatest difference in *IS* – values exists for the sites B and C, and that in *QS* for the site E, for which our calculations yielded 0.45 mm/s against 0 mm/s as obtained in [82]. This comparison of the data clearly shows that the proper analysis of the Mössbauer spectra recorded on σ–FeCr alloys cannot be in practice correctly done without supporting theoretical calculations as those outlined in [79].

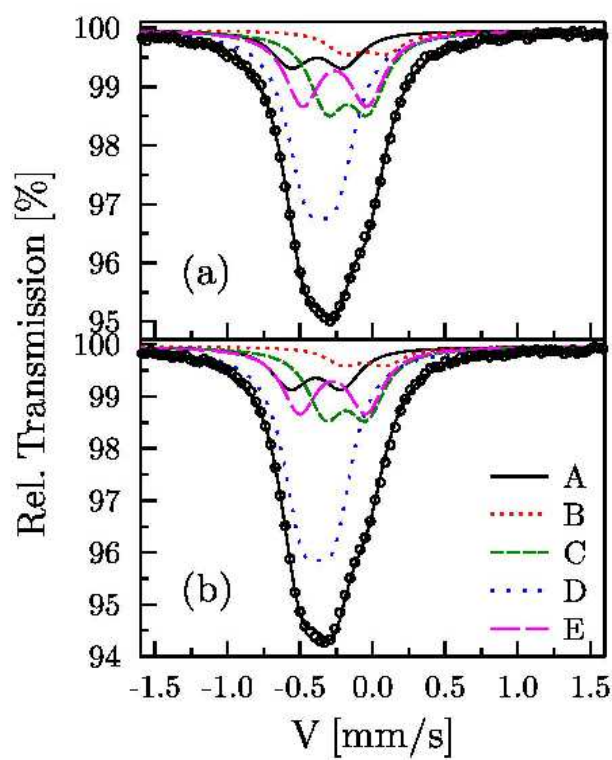

**Fig. 9.3**
*$^{57}$Fe-site Mössbauer spectra recorded at 295 K on two samples with (a) x = 37 and (b) x = 46. The best-fit spectrum and five subspectra are indicated by solid lines [redrawn from Ref. 83].*

Similar calculations have been recently performed for σ–FeV alloy system [83], where the σ-phase, as already mentioned, exists in a compositional range of ~33-60 at% V, giving thereby much better possibility for testing the calculations which, as described in Ref. 81, were successfully performed for two extreme values of *x* viz. 33.3 and 60, and they were next successfully used to analyze the Mössbauer spectra recorded in a paramagnetic state. For any intermediate *x*-value the analysis of the spectra was successfully performed using a linear approximation between the calculated extreme cases, which is displayed in Fig. 9.3.

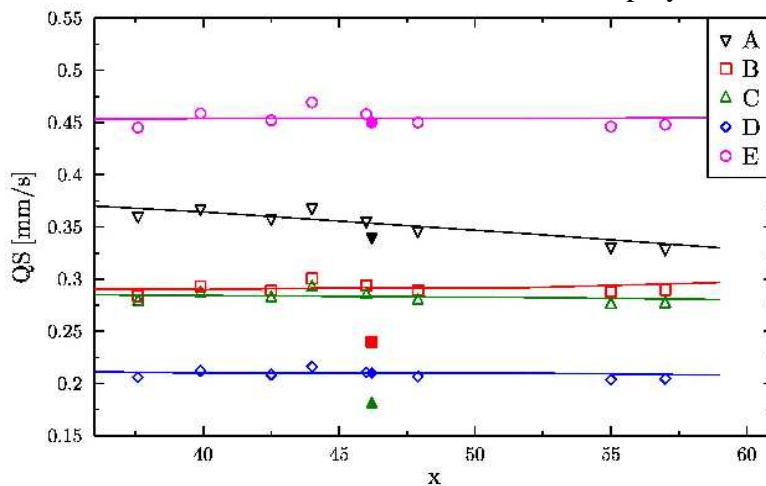

**Fig. 9.4**
*Quadrupole splitting, QS, as determined for each site and vanadium concentration, x, from the analysis of the measured spectra carried out with the protocol described in the text. QS-*

*values obtained previously for σ-FeCr compound are indicated as reversibly filled symbols for comparison [83].*

The calculations permitted determination of *QS*-values – see Fig. 9.4, as well as that of the average isomer shift, *<IS>*, as a function of vanadium concentration, *x*. The latter, that can be seen in Fig. 9.5, shows a very good agreement with the measured data.

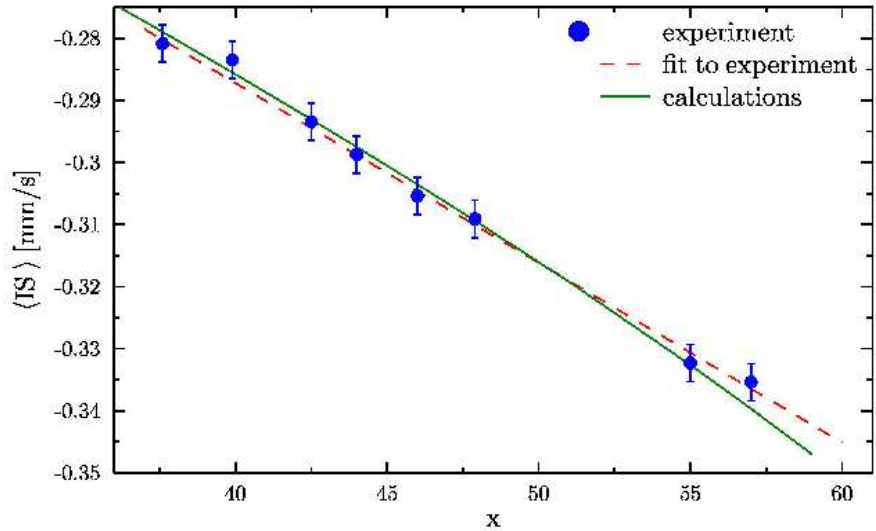

**Fig. 9.5**
*Average centre shift, < IS >, versus vanadium concentration, x, as measured (circles) and as calculated (solid line). The dashed line indicates the best-fit to the measured values [83].*

## 10. SUMMARY

The paper is devoted to the σ-phase in Fe-Cr and Fe-V alloy systems, the only ones where the magnetic properties of the phase are well evidenced. The emphasis is given on its physical properties as determined with the Mössbauer spectroscopy, though questions relevant to ranges of its occurrence in the temperature-composition plane, as well as those relevant to its identification and determination of its structure, site occupancy and the kinetics of transformation (precipitation and disappearance) are also reported. The latter issue is of particular interest due to an industrial importance of materials in which the σ-phase can precipitate causing thereby a drastic deterioration of their useful properties.

Concerning the physical properties, the ones determined experimentally as well as the corresponding ones obtained with theoretical calculations are presented. In particular, the lattice dynamics is addressed by describing measurements of the Debye temperature and those of the phonon density of states. The magnetic properties of the phase are represented by the results depicting the Curie temperature, the magnetic hyperfine field and the magnetic moment of Fe atoms.

Finally, theoretically obtained quantities that are pertinent to the electronic structure of the studied phase in Fe-Cr and Fe-V alloys are described.

## ACKNOWLEDGEMENTS
The Ministry of Science and Higher Education, Warsaw, is acknowledged for support.24